# Enhancing a Near-Term Quantum Accelerator's Instruction Set Architecture for Materials Science Applications

XIANG ZOU[1], SHAVINDRA P. PREMARATNE[1], M. ADRIAAN ROL[2],
SONIKA JOHRI[1], VIACHESLAV OSTROUKH[2], DAVID J. MICHALAK[1],
ROMAN CAUDILLO[1], JAMES S. CLARKE[1], LEONARDO DICARLO[2],
AND A. Y. MATSUURA[1]

[1]Components Research, Intel Corporation, Hillsboro, OR 97124 USA
[2]QuTech, Delft University of Technology, Delft, GA 2600, The Netherlands

Corresponding author: Xiang Zou (xiang.chris.zou@intel.com).

**ABSTRACT** Quantum computers with tens to hundreds of noisy qubits are being developed today. To be useful for real-world applications, we believe that these near-term systems cannot simply be scaled-down non-error-corrected versions of future fault-tolerant large-scale quantum computers. These near-term systems require specific architecture and design attributes to realize their full potential. To efficiently execute an algorithm, the quantum coprocessor must be designed to scale with respect to qubit number and to maximize useful computation within the qubits' decoherence bounds. In this work, we employ an application-system-qubit co-design methodology to architect a near-term quantum coprocessor. To support algorithms from the real-world application area of simulating the quantum dynamics of a material system, we design a (parameterized) arbitrary single-qubit rotation instruction and a two-qubit entangling controlled-Z instruction. We introduce *dynamic gate set* and *paging* mechanisms to implement the instructions. To evaluate the functionality and performance of these two instructions, we implement a two-qubit version of an algorithm to study a disorder-induced metal-insulator transition and run 60 random instances of it, each of which realizes one disorder configuration and contains 40 two-qubit instructions (or gates) and 104 single-qubit instructions. We observe the expected quantum dynamics of the time-evolution of this system.

## I. INTRODUCTION

Quantum computing is one of the most important nascent technologies today, with many recent advances in numbers of qubits [1]–[3]. However, there are many challenges to designing and building practical general-purpose quantum computers for real-world applications, and the mainstream use of quantum computers is expected to be at least a decade away [4]. Near-term quantum computers will likely have a few hundreds of noisy qubits without robust error-correction [5]. During this time, quantum devices will be limited to executing quantum circuits (a model for quantum computation) much smaller than ultimately possible with fault-tolerant quantum devices. Nevertheless, we expect these near-term devices to offer benefits to certain classes of quantum applications, such as quantum chemistry and materials design simulations.

In addition to more resilient qubits, architecture and design level solutions are required to reach the full potential of near-term devices. There has been active research focusing on general quantum computer architecture and on compilers for fault-tolerant large-scale quantum systems [6]–[12]. Recent breakthroughs in demonstrating programmable quantum computation on near-term devices [1] have allowed researchers to experiment using various basic quantum algorithms [13] and to develop circuits-to-qubits mapping methods [14] to meet architecture constraints. In contrast, our study enhances the architecture and design of a near-term quantum accelerator to speedup algorithm execution and

  

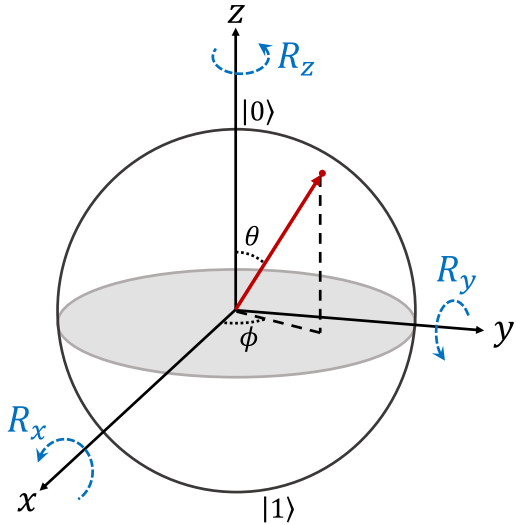

**FIGURE 1.** Representing the qubit state on the Bloch sphere. The qubit state as represented in Equation (1) is depicted on the surface of the Bloch sphere as a red dot, with $\theta$ and $\phi$ being the relevant angles. The poles of the sphere correspond to the states $|0\rangle$ and $|1\rangle$. Rotations around the positive $x$, $y$, and $z$ axes are represented by the dashed blue arrows. The filled gray circle represents the $xy$ plane.

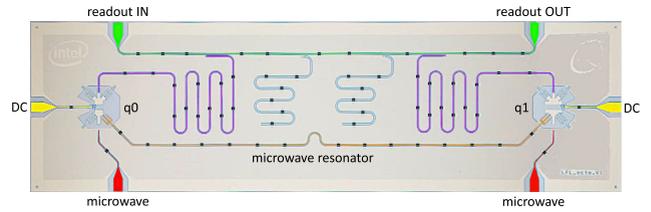

**FIGURE 2.** Octobox-2 transmon qubit chip schematic. Two transmon qubits $q0$ and $q1$ are coupled by a superconducting MW resonator. Each qubit has a dedicated MW drive port and a direct-current (dc) flux drive port. Both qubits are coupled to a readout feedline for measurement. The relaxation times for the qubits were $T_1^{(q0)} = 28$ $\mu$s and $T_1^{(q1)} = 22$ $\mu$s. The spin echo coherence times of the qubits measured at the system operating point were $T_2^{(q0)} = 4.2$ $\mu$s and $T_2^{(q1)} = 38$ $\mu$s.

highlights the usefulness of an algorithm-system-qubit co-design approach [15].

In this work, we determine necessary functionalities demanded from a near-term quantum coprocessor to run quantum algorithms to simulate materials systems. We identify that an instruction to perform an arbitrary single-qubit microwave (MW) operation and an instruction to execute a discrete two-qubit controlled-Z operation are two key functionalities required for efficient algorithm execution.

## II. BACKGROUND

The fundamental building blocks of quantum computers are quantum bits (qubits). Qubits are ideal two-level quantum systems that store quantum information. The state of a qubit is manipulated using quantum operations. Physically realizing systems that are suitable for consideration as qubits is a challenge [16], and numerous technologies are presently being considered to build ultimate fault-tolerant large-scale quantum computers.

### A. QUANTUM COMPUTING BASICS

The two levels of a qubit are denoted by $|0\rangle$ and $|1\rangle$, which are column vectors representing the two basis states. The state of a qubit denoted by $|\psi\rangle$ can be expressed generally as

$$|\psi\rangle = \cos\frac{\theta}{2}|0\rangle + e^{i\phi}\sin\frac{\theta}{2}|1\rangle \quad (1)$$

where the angle $\theta$ defines the relative weight of the superposition between states $|0\rangle$ and $|1\rangle$, and the angle $\phi$ defines the phase of the superposition state. The resulting two-dimensional continuous representation can be visualized on a unit sphere known as the Bloch sphere [17] (see Fig. 1). In contrast, the state of a classical bit is represented by only two discrete points. Though the qubit state can be a general superposition of the basis states, a qubit measurement will collapse the state to either $|0\rangle$ or $|1\rangle$ with probabilities $\cos^2\frac{\theta}{2}$ and $\sin^2\frac{\theta}{2}$, respectively. Quantum algorithms are crafted to take advantage of this probabilistic nature.

The quantum circuit model of computation is used to map a quantum algorithm on many qubits as a sequence of rudimentary components called *quantum gates*. Elementary quantum gates acting on a single qubit include rotation operations **R**$_x$, **R**$_y$, and **R**$_z$ (see Fig. 1). There is a class of elementary two-qubit gates that dedicate one qubit to act as a control for the other qubit to conditionally perform a single qubit gate. For example, the controlled-Z $q1$, $q0$ gate (or **cZ**) acts on the two qubits $q0$ and $q1$, and performs the Z operation on the target qubit $q1$ only when the control qubit $q0$ is in $|1\rangle$. The state of $q1$ is left unchanged if the control qubit is in $|0\rangle$. Note that this is a fully quantum operation, that does not involve a collapse and measurement of the control qubit. Such two-qubit gates are fundamental in quantum computing as they give rise to *quantum entanglement*, which along with *quantum superposition* are crucial components for the speedup expected from quantum computation. A comprehensive treatment on quantum computing theory can be found in [17].

### B. PHYSICAL REALIZATION OF QUBITS

Although research into the physical realization of qubits has been ongoing for a few decades [18], there is no consensus within the quantum computing community on the best implementation, due to advantages and disadvantages specific to each technology [19]–[23]. In this work, we use a superconducting system based on tunable transmon qubits [24]. A two-qubit Intel transmon chip similar to the one used in this work (which we refer to as Octobox-2) is shown in Fig. 2.

The state of a transmon qubit can be manipulated using precise control of the applied MW radiation at frequencies 3–9 GHz, and the interaction between qubits can be controlled by tuning the effective interqubit coupling [25]–[27]. Superconducting resonators are typically used as mediators for efficient qubit measurement [28], [29]. An extensive introduction to transmon qubits can be found in [24].

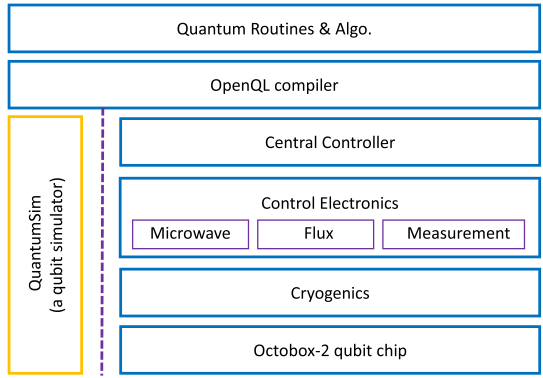

**FIGURE 3.** Simplified view of the bQI stack. The blue boxes label the main components with OpenQL compiler running on the host computer. QuantumInfinity also includes a density-matrix-based qubit simulator [33] to simulate quantum experiments using parameters extracted from transmon qubit characterization.

## C. QUANTUM COMPUTING SYSTEM

The TU Delft superconducting quantum computing stack (QuantumInfinity) along with Octobox-2 is schematically represented in Fig. 3. QuantumInfinity translates the quantum instructions into physical MW pulses and dc flux pulses in order to execute the quantum algorithms on qubits. The baseline QuantumInfinity (bQI) system stack supported the following native instructions (or basis gate set) for qubit characterization and benchmarking: $R_x(\pi)$, $R_x(0.5\pi)$, $R_y(\pi)$, $R_y(0.5\pi)$, and cZ $[q0, q1]$. Arbitrary waveform generators (AWGs) stored the pulses corresponding to these single-qubit and two-qubit operations.

The bQI stack employed OpenQL [30] as the compiler to generate executable code for the central controller (CC) [31]. The bQI system executed the single-qubit instructions similar to any programmable processing unit. The AWG stored a single pulse for each of the four instructions. Every instance of an instruction in a quantum program triggered the CC to command the AWG to drive the corresponding instruction's waveform via a codeword [32]. This method is distinctly different from the conventional technique of qubit control, where all required qubit operations are concatenated into a monolithic waveform and played out once.

## III. DISORDERED SPIN SYSTEM SIMULATION ALGORITHM

Simulation of materials that exhibit many-body physics effects, such as unusual electronic and magnetic properties, is often intractable on classical computers. In this work, we develop an algorithm to study the disorder-induced metallic and insulating dynamics exhibited in certain materials systems, which remains poorly understood [34], [35].

Consider a simplified model of a spin system described by the time-independent Hamiltonian for a one-dimensional nearest-neighbor Heisenberg model with random magnetic fields along both *x* and *z* directions [36] as follows:

$$\mathcal{H} = \sum_j \vec{\sigma}_j \cdot \vec{\sigma}_{j+1} + w \left( \sum_j h_j^x \sigma_j^x + \sum_j h_j^z \sigma_j^z \right) \quad (2)$$

where $\vec{\sigma}_j = \{\sigma_j^x, \sigma_j^y, \sigma_j^z\}$ denotes the Pauli vector for spin $j$ expressed in terms of Pauli matrices [17]. $h_j^x$ and $h_j^z$ characterize the strength of the disorder present in the system and are random variables chosen from a uniform distribution $[-1, 1]$. At small and large values of $w$, the system is said to be in the thermalized phase (metal) and the localized phase (insulator), respectively.

The dynamics of our system are described by

$$|\Psi(t)\rangle = \exp(-i\mathcal{H}t) |\Psi(0)\rangle \quad (3)$$

where $|\Psi(0)\rangle$ is the initial state of the system, and $|\Psi(t)\rangle$ is the state at time $t$. To obtain $|\Psi(t)\rangle$ on a quantum computer, we first decompose $\mathcal{H}$ into $\sum_{m=1}^{M} \mathcal{H}_m$, where each $\mathcal{H}_m$ corresponds to a supported gate. We use Trotterization [37] to divide the total evolution time $T$ into $N$ time steps of size $\tau$. With small enough $\tau$, the time evolution operator is well approximated by

$$\exp(-i\mathcal{H}t) \approx \left( \prod_{m=1}^{M} \exp(-i\mathcal{H}_m \tau) \right)^N. \quad (4)$$

Although the quantum algorithm is valid for an arbitrary number of spins, here we focus on the simplified two-spin version of the algorithm as our quantum chip supports two qubits ($q0$ and $q1$). We perform a one-to-one correspondence between spins and qubits, and simulate the spin dynamics by applying quantum gates in the form of control signals to the physical qubits.

Algorithm 1 lists the quantum gate sequence as described in the disorder-induced metal-insulator transition quantum algorithm. While several properties of the many-body system can be studied using this algorithm, we focus on the memory retention of $|\Psi(0)\rangle$ under time evolution. This property is described by the imbalance $\mathcal{I}$ defined as

$$\mathcal{I} = \mathcal{P}_0 - \mathcal{P}_1 \quad (5)$$

where $\mathcal{P}_j$ with $j \in \{0, 1\}$ is the probability of finding qubit $j$ in the $|1\rangle$ state. At time 0, $q0$ is in $|1\rangle$ state, $q1$ is in $|0\rangle$ state and, thus, $\mathcal{I} = 1$. The variation of imbalance is studied by averaging over many random disorder realizations.

## IV. ARCHITECTURE CONSIDERATIONS AND DESIGN

Previous quantum architecture investigations have focused on designing fault-tolerant large-scale quantum computers [6]–[9]. Here, we consider architecture and design to enable efficient execution of real-world quantum algorithms on near-term systems. Due to decoherence and gate errors, the available computation time on near-term devices is limited. This section discusses the quantum instruction set architecture enhancements to shorten circuit execution time so that

**Algorithm 1:** Instructions for Disorder-Induced Metal-Insulator Transition Algorithm on Two Spins.

for $k = 0$ to $N$ do
  reset $q0$
  reset $q1$
  $R_x(\pi)$ $q0$    // initialization: $q0 = |1\rangle$, $q1 = |0\rangle$
  for $m = 0$ to $k$ do
    cNOT $q1, q0$    // beginning of an **interval**
    **in parallel do**
      $R_z(2wh_0^z\tau)$ $q0$    // $q0$ random disorder
      $R_z(2\tau)$ $q1$
    **end do**
    $cR_x(4\tau)$ $q0, q1$
    cNOT $q1, q0$
    $R_z(2wh_1^z\tau)$ $q1$    // $q1$ random disorder
    **in parallel do**
      $R_x(2wh_0^x\tau)$ $q0$    // $q0$ random disorder
      $R_x(2wh_1^x\tau)$ $q1$    // $q1$ random disorder
    **end do**    // end of an **interval**
  end for
  q0mZ[k] ← measure $q0$    // q0mZ classical variable
  q1mZ[k] ← measure $q1$    // q1mZ classical variable
end for

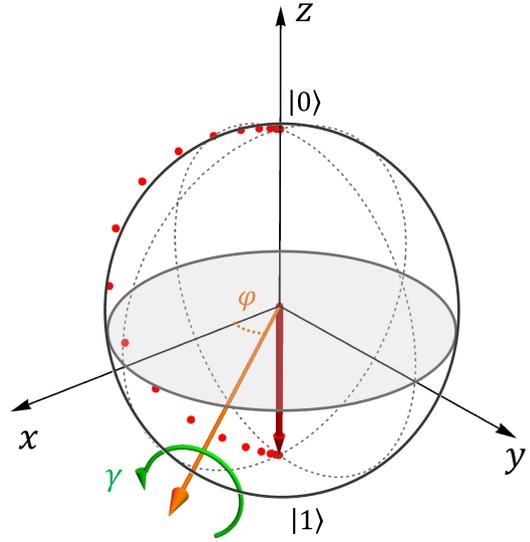

**FIGURE 4.** Evolution of the qubit state during a $R_{xy}(\varphi = 0.2\pi, \gamma = \pi)$ operation is shown. The axis defined by $\phi = 0.2\pi$ is represented by the orange arrow. The red arrow and the green circular arrows represent the final qubit state vector and the sense of rotation, respectively. The red dots on the Bloch sphere denote the most probable qubit state at each 1 ns timestep when the gate time is set to 20 ns.

the algorithm can perform more Trotter steps, i.e., coherently evolve over a longer time period. Here, we describe our modifications to the bQI stack, which resulted in the enhanced Quantum Infinity (eQI) system stack capable of fully instruction-based quantum processing.

### A. GATE DECOMPOSITION

Though Algorithm 1 is expressed with valid one-qubit and two-qubit gates, it is not possible to execute all these operations natively on the Octobox-2 qubit chip. We must first decompose non-native operations according to Algorithm 2 to obtain a set of operations that are executable on Octobox-2. Note that Decomposition 3 uses a new $R_{xy}$ instruction in the eQI system for arbitrary single-qubit rotations.

### B. SINGLE-QUBIT ROTATION $R_{xy}(\varphi, \gamma)$ INSTRUCTION

Our quantum algorithm specifies single-qubit rotations with random angles to model disorder in the spin system. These random angles are distinctly different from the standard rotation angles (i.e., $\pi$, $\pi/2$) supported by the bQI stack. In this study, native quantum gates are introduced in the eQI system stack to maximize the amount of useful computation within the limited coherence of the near-term devices. We add a single-pulse single-qubit rotation instruction

$$R_{xy}(\varphi, \gamma) = \begin{bmatrix} \cos(\gamma/2) & -ie^{-i\varphi}\sin(\gamma/2) \\ -ie^{i\varphi}\sin(\gamma/2) & \cos(\gamma/2) \end{bmatrix}. \quad (6)$$

This instruction applies a rotation of $\gamma$, around an axis oriented $\varphi$ measured from the $x$-axis on the $xy$ plane (see Fig. 4).

**Algorithm 2:** Gate Decompositions.

// **Decomposition 1**: cNOT $q0, q1$
$R_y(-0.5\pi)$ $q0$
cZ $q0, q1$
$R_y(0.5\pi)$ $q0$
// **Decomposition 2**: $cR_x(\alpha)$ $q0, q1$
cZ $q0, q1$
$R_x(-0.5\alpha)$ $q0$
cZ $q0, q1$
$R_x(0.5\alpha)$ $q0$
// **Decomposition 3**: $R_z(\beta)$ $q1$
$R_{xy}(\varphi = 0.5\pi, \gamma = \pi)$ $q1$
$R_{xy}(\varphi = 0.5\beta - 0.5\pi, \gamma = \pi)$ $q1$

Both $\varphi$ and $\gamma$ are continuous variables. Note that $R_x(\gamma) \equiv R_{xy}(\varphi = 0, \gamma)$ and $R_y(\gamma) \equiv R_{xy}(\varphi = 0.5\pi, \gamma)$. Arbitrary rotations are typically physically implemented simply by adjusting the phase of the MW pulses. The techniques introduced in the eQI system to support arbitrary rotation instructions are shown in Fig. 5.

$R_{xy}(\varphi, \gamma)$ is an instruction that effectively specifies an infinite number of rotation operations. This is challenging to implement, since the AWG has limited memory (to store waveforms) and limited codeword space (to address individual gates). We introduce and implement three mechanisms on the eQI stack to address these hardware constraints: *quantum operation specification* (QOS), *dynamic gate set* (DGS), and *paging* (PG).

The QOS maintained on the host computer defines the extended discrete set of quantum operations supported by the eQI system. It contains the list of angles and axes permitted for single-qubit rotations, and the signal pulses to perform each rotation on the physical qubits. DGS preprocesses the quantum program each time we run it, and identifies all different single-qubit rotations used in the quantum program. If any single-qubit rotation is new with respect to QOS, DGS augments QOS and generates the appropriate MW control pulses.

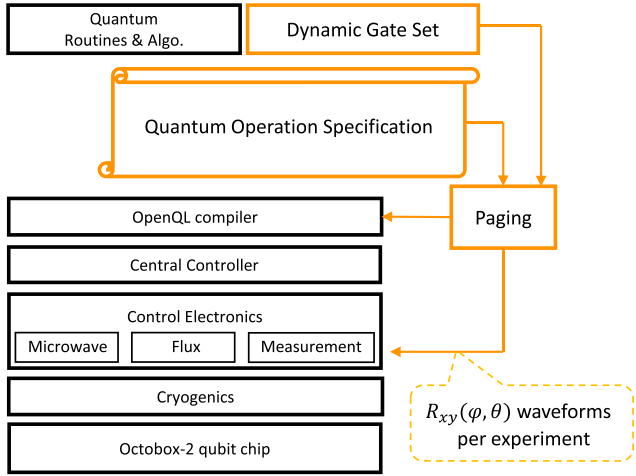

**FIGURE 5.** Simplified view of the eQI stack. The quantum operation specification, a dynamic gate set, and paging are used to implement single-qubit arbitary-rotation instructions.

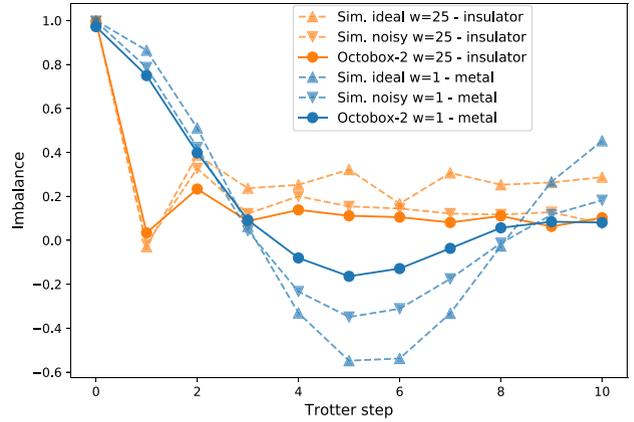

**FIGURE 6.** Results from Trotterization of disorder-induced metal-insulator transition averaged over 60 disorder realizations. Filled circles on solid lines represent data from the experiment on Octobox-2 qubit chip and filled triangle on dashed lines represent results from simulations on QuantumSim [33] using both ideal and realistic noisy qubits. The ideal simulation shows that at large disorder, the Imbalance converges to a constant nonzero value while in the low-disorder system, the Imbalance keeps oscillating up to the longest time simulated.

PG manages a rotation-to-codeword table (RCT), which tracks the rotations stored in the electronics memory, and loads all MW pulses required by the program into the electronics according to the RCT. PG implements the following scheme to reduce the waveform load time. First, PG determines the missing list (MLST), i.e., the list of rotations required by the program but not in the electronics, and the dumping list (DLST), i.e., the list of rotations not used by the program but stored in the electronics. The number of elements in the MLST is not more than the DLST. Second, PG randomly selects RCT entries in the DLST and replaces them by the MLST. Third, PG loads the MW pulses corresponding to the updated RCT entries into electronics.

The OpenQL compiler consumes the dynamically generated QOS and RCT, and creates a CC executable. By ensuring the number of unique rotations in an experiment does not exceed the available memory or the codeword space, the QOS, DGS, and PG allowed us to efficiently support the single-qubit arbitrary rotation instructions required by the algorithm.

### C. TWO-QUBIT cZ INSTRUCTION

Supporting the discrete cZ instruction in QuantumInfinity was particularly challenging due to the long-timescale hysteresis effects present on the control pulse path in the hardware (e.g., bias tees and low-pass filters). In previous qubit experiments, the distortion resulting from hysteresis was addressed by forming a single precompensated monolithic waveform containing all cZ operations, appropriately spaced to synchronize with single-qubit operations. The first cZ instance is interpreted by the quantum coprocessor to trigger the single concatenated waveform, while subsequent instances are processed as dummy instructions.

The single-trigger instruction method, as described above, works well for most short-duration qubit experiments and characterization. However, it is intractable for the real-world algorithms that use many qubits or use feedback based on qubit measurements to determine subsequent quantum operations. The problems arising from the single-trigger multi cZ instruction was addressed by implementing a robust physical cZ operation using the Net-Zero method [38], and by activating real-time digital filters of the AWG [39]. These enhancements minimized the hysteresis effects on flux pulses and, thus, enabled the repeatable cZ instruction.

## V. ALGORITHM RESULTS ON OCTOBOX-2 CHIP

A single realization of the decomposed quantum algorithm as run on the Octobox-2 chip is shown in Algorithm 3. All of the operations represent the native gates of eQI, which now include $R_{xy}$ and cZ. Note that we take advantage of the fact that this algorithm only involves two disorder fields along $x$- and $z$-axis, respectively, and rotate the single-qubit basis so that $z$ becomes $y$, i.e., turning all disorder-specific $R_z(2wh_j^z\tau)$ into $R_y(2wh_j^y\tau)$ with $j \in \{0, 1\}$. This frame rotation eliminates two $R_z$ decompositions, while keeping the physics intact.

We run the algorithm for $w = 1$ and $w = 25$, with the Trotter step size $\tau = 0.04\pi$. $h_0^x, h_0^y, h_1^x, h_1^y$ are assigned random numbers independently sampled from $[-1, 1]$. Each assignment to an input set $\{h_0^x, h_0^y, h_1^x, h_1^y\}$ corresponds to a single disorder realization. The experiment on Octobox-2 runs the algorithm over $N = 10$ Trotter steps with a total of

**Algorithm 3:** Quantum Coprocessor Instructions for a Single Realization of the Disorder-Induced Metal-Insulator Transition Algorithm on Two Spins.

**Input:** $w, h_0^x, h_0^y, h_1^x, h_1^y$
  **for** $k = 0$ to $N$ **do**
    // prologue
    **in parallel do**
      reset $q0$
      reset $q1$
    **end do**
    **in parallel do**
      $\mathsf{R}_{xy}(0, 0.5\pi)\ q0$
      $\mathsf{R}_{xy}(0, -0.5\pi)\ q1$
    **end do**
    **for** $m = 0$ to $k$ **do**
      // main body
      **in parallel do**
        $\mathsf{R}_{xy}(0.5\pi, 0.08wh_0^y\pi)\ q0$
        $\mathsf{R}_{xy}(0.5\pi, -0.5\pi)\ q1$
      **end do**
      cZ $q1, q0$
      $\mathsf{R}_{xy}(0.5\pi, -0.5\pi)\ q1$
      $\mathsf{R}_{xy}(-0.46\pi, \pi)\ q1$
      cZ $q0, q1$
      $\mathsf{R}_{xy}(0, -0.08\pi)\ q0$
      cZ $q0, q1$
      **in parallel do**
        $\mathsf{R}_{xy}(0, 0.08\pi)\ q0$
        $\mathsf{R}_{xy}(0.5\pi, -0.5\pi)\ q1$
      **end do**
      cZ $q1, q0$
      $\mathsf{R}_{xy}(0.5\pi, 0.5\pi + 0.08wh_1^y\pi)\ q1$
      **in parallel do**
        $\mathsf{R}_{xy}(0, 0.08wh_0^x\pi)\ q0$
        $\mathsf{R}_{xy}(0, 0.08wh_1^x\pi)\ q1$
      **end do**
    **end for**
    // epilogue
    **in parallel do**
      $\mathsf{R}_{xy}(0, 0.5\pi)\ q0$
      $\mathsf{R}_{xy}(0, 0.5\pi)\ q1$
    **end do**
    **in parallel do**
      q0mZ$[k] \leftarrow$ measure $q0$
      q1mZ$[k] \leftarrow$ measure $q1$
    **end do**
  **end for**
**Output:** q0mZ, q1mZ

60 disorder realizations. For each input set, the algorithm is repeated $n_{\text{avg}}$ times at each Trotter step. The values of q0mZ and q1mZ are then averaged over the $n_{\text{avg}}$ runs to estimate the probabilities $\mathcal{P}_0$ and $\mathcal{P}_1$ at each Trotter step. The imbalance $\mathcal{I}$ at each Trotter step is then calculated by averaging the results over all 60 disorder realizations for a given value of $w$.

To compare with the experimental results, we run the quantum algorithm on a density-matrix simulator on a classical computer as well [33]. We model the qubit system both in the absence and presence of noise and decoherence, and simulate Algorithm 3 as in the experiment. The results from the simulator and experiment are shown in Fig. 6.

For both $w$ values, Octobox-2 data exhibits behavior similar to the simulations. In particular, there is good agreement between the data and simulations for the first few Trotter steps. The Octobox-2 data clearly demonstrate the difference in the dynamical behavior between $w = 1$ and $w = 25$, indicating the metallic and insulating phases, respectively. We attribute most of the disagreement observed between the data and simulations to operating $q0$ away from its sweet spot, leading to increased dephasing and suboptimal performance during idle and single-qubit gates.

## VI. CONCLUSION

In this work, we employ an algorithm-system-qubit co-design methodology to architect and design a near-term quantum coprocessor for the real-world application area of simulating quantum dynamics in a materials science system. We develop a small algorithmic workload to study the dynamics of a disordered, interacting system and implement specific functionalities (discrete $\mathsf{R}_{xy}$ and cZ instructions) that a quantum coprocessor must support for efficient quantum algorithm execution.

Recent related experiments reported in the literature either have only a few entangling gates [40]–[42] or include only a few arbitrary angles [43]. Our quantum algorithm presented unique architectural and design challenges in requiring a large number of single-qubit arbitrary rotations and two-qubit entangling gates. We implemented the two-qubit version of the algorithm with 60 different random disorder realizations, each of which contained 40 two-qubit instructions and 104 single-qubit instructions. The results validated the capability of our enhanced quantum computing system to run small-scale algorithms targeting real-world applications.

We recognize that the results from state-of-the-art, few-qubit devices are still limited in accuracy. The aim of this work is to implement necessary architectural features to maximize the potential of near-term systems. By utilizing larger quantum computers, we believe that there is potential to gain more accurate information about such materials systems, even with noisy qubits.

### ACKNOWLEDGMENT
The authors would like to thank Ramiro Sagastizabal for insightful discussions on utilization of virtual $R_z$ gates.